\documentclass[aps,pra,twocolumn,superscriptaddress]{revtex4}

\pdfoutput=1

\usepackage{amsmath,graphicx,amssymb,braket,xcolor,subfigure,upgreek,bbold,bm}
\usepackage[colorlinks, linkcolor=blue, citecolor=blue, urlcolor=blue, breaklinks=true]{hyperref}
\usepackage{verbatim}
\usepackage{float}
\usepackage{esint}
\usepackage{bm}
\usepackage{bbold}
\newcommand{\mi}{{\rm i}}
\newcommand{\me}{{\rm e}}

\begin{document}

\title{Adiabatic preparation of entangled, magnetically ordered states \\with cold bosons in optical lattices}

\author{Araceli Venegas-Gomez}
\affiliation{Department of Physics and SUPA, University of Strathclyde, Glasgow G4 0NG, UK}
\author{Johannes Schachenmayer}
\affiliation{CNRS, IPCMS (UMR 7504), ISIS (UMR 7006), and Universit\'{e} de Strasbourg, 67000 Strasbourg, France}
\author{Anton S. Buyskikh}
\affiliation{Riverlane, Cambridge CB2 3BZ, UK}
\author{Wolfgang Ketterle}
\affiliation{Harvard-MIT Center for Ultracold Atoms, Cambridge, Massachusetts 02138, USA}
\affiliation{Department of Physics, Massachusetts Institute of Technology, Cambridge, Massachusetts 02139, USA}
\author{Maria Luisa Chiofalo}
\affiliation{Dipartimento di Fisica “Enrico Fermi”, Universit\'{a} di Pisa and INFN, Largo B. Pontecorvo 3, I-56127 Pisa, Italy}
\author{Andrew J. Daley}
\affiliation{Department of Physics and SUPA, University of Strathclyde, Glasgow G4 0NG, UK}

\date{\today}


\begin{abstract}

We analyze a scheme for preparation of magnetically ordered states of two-component bosonic atoms in optical lattices. We compute the dynamics during adiabatic and optimized time-dependent ramps to produce ground states of effective spin Hamiltonians, and determine the robustness to decoherence for realistic experimental system sizes and timescales. Ramping parameters near a phase transition point in both effective spin-1/2 and spin-1 models produces entangled spin-symmetric states that have potential future applications in quantum enhanced measurement. The preparation of these states and their robustness to decoherence is quantified by computing the Quantum Fisher Information of final states. We identify that the generation of useful entanglement should in general be more robust to heating than it would be implied by the state fidelity, with corresponding implications for practical applications.
  \end{abstract}

\maketitle

\section{Introduction}

Over the past few years, the level of control available over ultracold atomic gases in optical potentials has opened many opportunities for the study of many-body time-dependent dynamics \cite{JakschZoller2005,Lewenstein2007,Bloch2008,Bloch2012,Gross2017}. One specific application of this is the possibility to use such dynamics to prepare and explore interesting quantum states. Such studies are not only restricted to better understanding strongly correlated systems, but have been applied also to quantum metrology, especially reducing systematic and statistical errors in leading platforms for optical atomic clocks \cite{Campbell90,Hutson2019}. 

The idea of exploiting many-particles correlations to enhance the phase sensitivity up to the Heisenberg limit has been proposed in the context of Mach-Zehnder  matter-wave interferometers, conceived in the form of a double well with Bose-Einstein Condensates as input states ~\cite{Jaaskelainen2004,Pezze2005}. In this geometry, metrologically useful squeezing has been experimentally demonstrated~\cite{Esteve2008} with a number of variants including the use of integrated atom chips~\cite{Berrada2013} and the tuning of atomic interactions into the attractive regime~\cite{Trenkwalder2016}. 
There have been significant recent experimental developments particularly in controlling the dynamics of spin systems formed from two-component bosonic gases in optical lattices 
\cite{Dimitrova2020,Amato2019}. The corresponding spin models arise from the two-component Bose-Hubbard model in the limit of strong interactions, and depending on the average filling factor of the lattice, models with different effective spin can be realized \cite{Kuklov2003,Altman2003}. At the same time there has been a lot of interest in questions of where atoms in optical lattices can produce quantum states with substantial entanglement in a form that could be useful for quantum enhanced metrology. In particular, several recent works have discussed the availabilty of such states near certain phase transition boundaries \cite{Liu_2013,YE2016151,Sinatra2018,Frerot2018,PhysRevX.8.021022,PhysRevA.99.043618,Frerot2019,PhysRevLett.123.060406,Hauke2016,Gabbrielli_2019,Gabbrielli2018}.
The usefulness of this entanglement for metrology is generally quantified via the Quantum Fisher Information (QFI) \cite{T_th_2014,Pezze2014,PhysRevLett.72.3439, Paris2009,Helstrom,RevModPhys.90.035005}, which makes it possible to characterize the potential of parameter estimation with a particular initial state to beat the Standard Quantum Limit (SQL) or shot noise limit of a scaling as $1/\sqrt{M}$, where $M$ is the number of particles in the system. 
 
Motivated by this, here we investigate schemes to time-dependently prepare states near phase transition boundaries in the effective spin models derived from the Bose-Hubbard model for two-species of atoms in an optical lattice. We show how these states completely symmetric in spin have a strongly enhanced QFI, and then compute and analyze the corresponding preparation dynamics using numerical techniques with Matrix Product States (MPS) \cite{White2004, SchollwockDaleyVidal,VerstraeteCiracMurg2008,Schollwock2011}. These are based on adiabatic state preparation techniques, beginning from a gapped initial state that can be prepared with low entropy, in a regime where the Hamiltonian is gapped. This is then followed by a slow ramp of Hamiltonian parameters to produce the required state \cite{Rabl2003,Rey2007,Ho2007,Soerensen2010,Schachenmayer2015,Trebst2006,Kantian2010,Lubasch2011}. We study the robustness of these ramps in the presence of dissipation, for typical experimental system sizes and timescales.  

The rest of this article is organized as follows: in Sec.~\ref{sec:QFI-GS} we introduce the spin models, as well as the QFI, and discuss entanglement properties of the magnetically ordered ground states. In Sec.~\ref{sec:ASP} we analyze how to prepare these ground states via a sequence of adiabatic ramps. In Sec.~\ref{sec:Dissipation} we then study the stability of these protocols when adding dissipation into the system, before providing a summary and outlook in Sec.~\ref{sec:Summary}. 

\section{Characterizing the Ground State}
\label{sec:QFI-GS}

In this section, we describe the two effective spin models arising from the Bose-Hubbard model for two species of bosons in an optical lattice, and characterize the ground states of these models near their isotropic points in terms of their QFI. 

\subsection{Spin Models}
\label{sec:Spin Models}

The model that is studied in this work is the Bose-Hubbard Model for two components ($A$ and $B$, corresponding to atomic or spin species) of bosonic atoms in an optical lattice, where atoms are confined to the lowest Bloch band \cite{Duan2003,Kuklov2003}, with Hamiltonian:
\begin{equation}
\begin{split}
\hat{\mathcal{H}}= & -\zeta{\sum_{\braket{i,j}}} (\hat{a}_i^\dagger \hat{a}_j+\hat{b}_i^\dagger \hat{b}_j)+U_{AB}{\sum_j}\hat{a}_j^\dagger \hat{a}_j\hat{b}_j^\dagger\hat{b}_j \\
& +\frac{U_A}{2}{\sum_j}\hat{a}_j^\dagger\hat{a}_j^\dagger\hat{a}_j \hat{a}_j +\frac{U_B}{2}{\sum_j}\hat{b}_j^\dagger \hat{b}_j^\dagger\hat{b}_j \hat{b}_j. 
\end{split}
\label{eq:2bosons_H_Diagram}
\end{equation}
Here, $\zeta$ designates the tunneling energies for the species $A$ and $B$, $\hat{a}_i^\dagger,\hat{a}_i$, $\hat{b}_i^\dagger,\hat{b}_i$  are the bosonic creation and annihilation operators on site $i$, respectively. $U_{A},U_{B}$ are the intra-species interactions, and $U_{AB}$ the inter-species on-site interaction. The notation $\braket{i,j}$ denotes a sum over nearest-neighbour sites.
The ground-state phase diagram was analysed by theoretical and computation methods e.g. computed in \cite{Altman2003}.

For strong interactions ($U_{AB},U_{A},U_{B}\gg \zeta$) the atoms become localized on individual lattice sites for integer filling. If the atoms are a mixture of two spin-components, this is effectively a system of spins which are held to a lattice, interacting by superexchange.

In the Mott Insulator regime where $U_{A}=U_{B}=U\gg\zeta$, the Hamiltonian can be simplified using second order perturbation theory in the tunneling \cite{Kuklov2003}. For a particular integer occupation $n$ on every site one can derive an effective low-energy Hamiltonian. 

In the case of $n=1$ of one particle per site, by applying second order perturbation theory on the tunneling, the resulting effective Hamiltonian is a spin-1/2 $XXZ$ Heisenberg Model. In this case, spin $\ket\uparrow$ represents the atomic species $A$ and spin $\ket\downarrow$ represents the atomic species $B$. The effective Hamiltonian takes the form:
\begin{equation}\label{eq:Spin12-H}
\hat{\mathcal{H}}_{\text{SP1/2}}= - J{\sum_{\braket{i,j}}}({\hat{s}^x_i}{\hat{s}^x_j}+{\hat{s}^y_i}{\hat{s}^y_j})-(J-\Delta){\sum_{\braket{i,j}}}{\hat{s}^z_i}{\hat{s}^z_j}
\end{equation}
with $ J=4{\zeta^2}/{U_{AB}}$ the superexchange, and $\Delta=8(\zeta^2/U_{AB})-8(\zeta^2/{U})$ represents the coupling anisotropy in this $XXZ$ Heisenberg Model, with $\hat{\bm{s}}_i=({\hat{s}^x_i,\hat{s}^y_i,\hat{s}^z_i})$ the three spin-1/2 operators.

Having an integer filling with $n=2$ atoms per site, and considering the case with same number of atoms of each species ($ {n}_{A}= {n}_{B}$), the effective Hamiltonian results in an anisotropic spin-1 Heisenberg Model with:
\begin{equation}\label{eq:Spin1-H}
\hat{\mathcal{H}}_{\text{SP1}}= - J{\displaystyle\sum_{\braket{j,l}}}{\hat{\textbf{S}}_{j}}{\hat{\textbf{S}}_{l}}+u{\displaystyle\sum_{l}}{(\hat{S}^{z}_{l})^{2}}
\end{equation}
where $ u=U - U_{AB} $, $J=4{\zeta^2}/U_{AB}$, and $({\hat{S}^x_i,\hat{S}^y_i,\hat{S}^z_i})$ is a vector of the three spin-1 operators.

In what follows, we compute ground states and dynamics (including dissipative dynamics) using methods based on MPS \cite{Schollwock2011,White2004,SchollwockDaleyVidal,VerstraeteCiracMurg2008}. We checked convergence of the calculations in the bond dimension, and the corresponding values are indicated in the figure captions.

We note that for computational purposes, we calculate all ground state properties and time evolution for state preparation for these models in 1D. We expect that the general principles of the highly entangled states will hold as discussed in higher dimensions, but it is much more difficult to compute the timescales required for successful adiabatic state preparation. These different dimensionalities would be easily accessible in experiments with cold atoms in optical lattices, and the effects of dimensionality would be very interesting to explore in that context. 

\subsection{Useful entanglement for metrology}
\label{sec:metrology}

Arising from studies of parameter estimation in metrology \cite{PhysRevLett.72.3439, Paris2009,T_th_2014,Pezze2014}, the QFI determines the optimal sensitivity (with the right measurement choice) of a state to a given transformation $\hat{G}$. It corresponds to the upper bound of the Fisher Information over all possible generalized quantum mechanical measurements~\cite{PhysRevLett.72.3439}, providing a tool to measure many-body entanglement useful for metrological purposes~\cite{Pezze2009,Hyllus2012,Toth2012}.

The more general concept of Fisher information $\mathcal{I}$ arises from the context of Quantum Estimation Theory \cite{PhysRevLett.72.3439, Paris2009}. Here we consider its specific application to quantum metrology in a setting where for a given state, described by a density operator $\hat{\rho}$, we perform a transformation of the state generated by $\hat{G}$, so that $\hat{\rho}(\theta) = \me^{-\textit{i}\theta\hat{G}}\hat{\rho}{\me}^{\textit{i}\theta\hat{G}}$ with a phase shift $\theta$ that we would like to measure. We can define an estimator $\theta_{\text{est}}$ based on a particular choice of measurement (generally a projective measurement in some basis), and assume we perform $m$ measurements. The resulting error associated with this estimator is subject to the Cram\'er-Rao bound \cite{T_th_2014,Pezze2014,Pezze2018}:
\begin{equation}
\Delta\theta_{\text{est}}\geq1/\sqrt{m\mathcal{I}}
\end{equation}
where $\Delta\theta_{\text{est}}$ is the variance of our estimator. 
By maximizing $\mathcal{I}$ over the set of all possible choices of measurement we obtain the Quantum Fisher Information $\mathcal{I_{Q}}$.
Then, we can restate the Cram\'er-Rao bound as:
\begin{equation}
\Delta\theta\geq1/\sqrt{m\mathcal{I}}\geq1/\sqrt{m{\mathcal{I_{Q}}}}.
\end{equation}

For product states across the atoms, the sensitivity of phase estimation is restricted to the shot noise limit, with $\Delta\theta\geq1/\sqrt{mM}$, being $M$ the total number of spins. However, these can be overcome by the introduction of entanglement in the system up to the Heisenberg scaling with $\Delta\theta\geq1/\sqrt{m}M$, corresponding to a chance in the scaling of $\mathcal{I_{Q}}$, from $\mathcal{I_{Q}}\propto{M}$, in the shot noise limit, to $\mathcal{I_{Q}}\propto{M}^2$ for the Heisenberg limit. We note that when $G$ can be written as the sum over the same operator $\hat G_l$ acting on each local spin $l$, $\hat G=\sum_l \hat G_l /2$ then the constant of proportionality in each limit is the square of the difference between the extreme eigenvalues of $\hat G_l$.

In the models described in this work, where we have many spins, in order to describe an ensemble of $M$ spins, we can introduce the collective spin vector $\hat{\textbf{J}}=\lbrace{\hat{J}_x,\hat{J}_y,\hat{J}_z}\rbrace$, where for spin-1/2 \cite{Pezze2018}
\begin{equation}
\hat{J}_\mu=\sum^M_{l=1}{\hat{s}^{(\mu)}_{l}},
\end{equation}
with $\hat{s}_{l}$ the spin operator for the particle $l$ and $\mu=x,y,z$ axis.

\begin{figure}[tb]
\centering
\includegraphics[width=6cm]{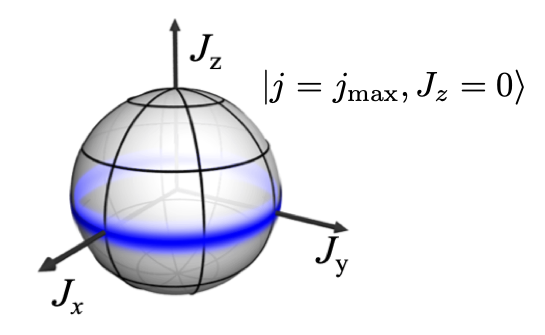}
\caption{Near phase transition points, we expect significant entanglement in the ground state. Here, near the isotropic point in both spin-1/2 and spin-1 models, the limiting state corresponds to a completely symmetric spin state (or Dicke state), with the maximum possible total angular momentum, and a z-component $J_z=0$, which can be depicted as a distribution of $\vec J$ values around the equator of the Bloch sphere (for the spin-1/2 case). Using an appropriate measurement scheme, this is a potential starting point for quantum enhanced measurement and sensing beyond the shot noise limit.}
\label{fig:Bloch}
\end{figure}

The target state (i.e., the ground state of the Hamiltonian in the limit of zero anisotropy) should be in the completely symmetric spin subspace (based on the symmetry of the completely isotropic Heisenberg Hamiltonian), but should also have $\langle \hat J_{z}\rangle = 0$, in order to minimise the energy for an infinitesimal anisotropy. This state lies around the equator of the enlarged Bloch Sphere as depicted in (Fig.~\ref{fig:Bloch}). An appropriate transformation for the QFI would be a coherent rotation away from the xy-plane, for which the generator is a collective spin operator:
\begin{equation}
\hat{G}=\hat{J}_x=\sum^M_{l=1}{\hat{s}^{(\mu)}_x}.
\end{equation}
This would correspond metrologically to a precision measurement of a magnetic field in the x-direction (note that any other direction in the x-y plane may be chosen). We choose to use this generator to characterise the QFI for the remainder of this work. 

With pure states we only need to compute the variance of the expectation value of the operator of our given transformation \cite{Pezze2014,T_th_2014}, and in our case
\begin{equation}\label{eq:QFIa}
\mathcal{I_{Q}}= 4\braket{(\Delta\hat{G})^2}= 4\braket{(\Delta\hat{J}_x)^2},
\end{equation}
with $\braket{(\Delta \hat O)^2}=\braket{\hat O^2}-\braket{\hat O}^2$ for any operator $\hat O$. That is, we compute the QFI for the spin-1/2 models as
\begin{equation}\label{eq:QFI}
\mathcal{I_{Q}}= 4(\braket{\hat{J}^{2}_x}-\braket{\hat{J}_{x}}^{2})=4\bigg(\sum_{l,l'}{\braket{\hat{s}^{x}_{l}\hat{s}^{x}_{l'}}}-\sum_{l}\braket{\hat{s}^{x}_l}^{2}\bigg).
\end{equation}
As pointed out in Refs.~\cite{Pezze2009,Hyllus2012,Toth2012}, in this sense we can use $\mathcal{I_{Q}}$ as a tool to quantify many-body entanglement. In particular, if $I_{Q}/M$ is larger than a given integer $z$, it implies that at least $z$-body entanglement is present. As noted above, it also provides a bound for the error in parameter estimation from an operation induced by the generator (in this case $\hat{J}_x$) \cite{Hyllus2012,Toth2012,T_th_2014, Pezze2014}. For spin-1 we compute the QFI analogously, using the spin-1 operators, $\hat{S}^x_{l}$. 

In this work, we will compute the QFI for different values of the anisotropies $\Delta/J,u/J$ in the ground states of the spin-1/2 and spin-1 model, respectively, as well as time-dependence below. We also study how $\mathcal{I_{Q}}$ scales with the system size $M$. The maximum  maximum QFI possible for our chosen generator $\hat G=\hat J_x$ and a given system size $M$ is $M^2$. The expression for the QFI in the target state, which is the maximum QFI we expect from our state preparation protocol, is given by
\begin{equation}\label{eq:QFI_max}
\mathcal{I_{Q}}_{\text{max}}= 4\bigg(\frac{J_{\rm max} (J_{\rm max}+1)}{2}\bigg),
\end{equation}
where $J_{\rm max}$ is the maximum total spin for the system ($J_{\rm max}=M/2$ for spin-1/2 and $J_{\rm max}=M$ for spin-1), so for the two models, $\mathcal{I_{Q}}_{\text{max}}^{\rm{SP1/2}}= M(\frac{M}{2}+1)$ and 
$\mathcal{I_{Q}}_{\text{max}}^{\rm{SP1}}= 2M(M+1)$.

We will address the question how close we are to $\mathcal{I_{Q}}_{\text{max}}$ in each model, and whether we have indeed useful entanglement, which could be a basis for quantum metrology beyond the shot noise limit (i.e., how close we are to a scaling $I_Q\propto M^2$, giving Heisenberg limited scaling as opposed to $I_Q\propto M$, which gives the standard quantum limit).

\begin{figure}[tb]
    \centering
    \includegraphics[width=\columnwidth]{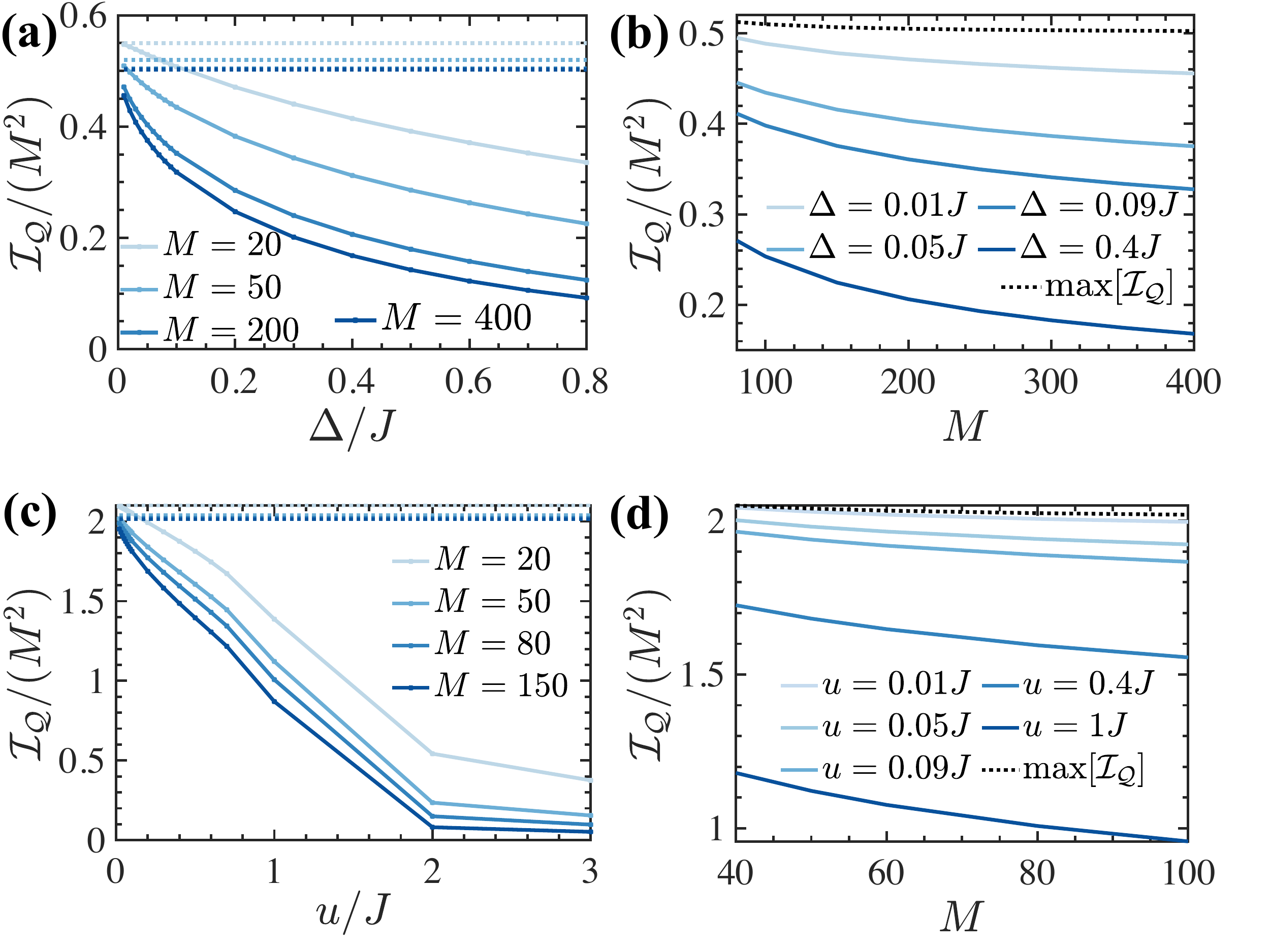}
    \caption{(a/c) Quantum Fisher Information $\mathcal{I_{Q}}$ for the ground state of the Hamiltonian for (a/b) the spin-1/2 model and (c/d) the spin-1 model. In (a) and (c) this is shown as a function of the anisotropy $\Delta/J,u/J$, and in (b) and (d) as a function of system size. As the anisotropy decreases the QFI approaches its theoretical maximum value \eqref{eq:QFI_max} for the observable $\hat J_x$ (dotted lines). At the same time, the scaling with system size approaches the Heisenberg limit, i.e., $\alpha\longrightarrow2$ and $\mathcal{I_{Q}}\sim{M}^2$. We note that the data points are specified by the markers, and the lines are added as a guide to the eye.  The results are obtained via MPS methods with bond dimension $D\leq256$, with open boundary conditions.}
    \label{fig:QFI_GS}
    \end{figure}

For both models the QFI has been calculated for different anisotropy values ($\Delta/J$, $u/J$ in each case), in Fig.~\ref{fig:QFI_GS}. We have also evaluated the scaling with system size by fitting the values of $\mathcal{I_{Q}}$ for different system sizes to a curve of the form 
\begin{equation}
\mathcal{I_{Q}}\propto{A}\cdot M^{\alpha}+\text{C},
\end{equation}
with $\alpha$ the scaling factor and $\text{A},\text{C}$ constants.

In the regime where we have larger nearest-neighbour interactions (smaller anisotropy), we see a larger value of $\mathcal{I_{Q}}$, so that the uncertainty in the parameter estimate approaches Heisenberg scaling in the limit of anisotropies $u/J,\Delta/J\longrightarrow 0$. We can see this, as the lines for different $M$ approach each other, and the dependence on $M$ of $I_{Q}/M^2$ becomes very weak. At larger anisotropies, we found a smaller Quantum Fisher information.
For the spin-1/2 system we have an XY-ferromagnet for all $\Delta/J>0$, and $\mathcal{I_{Q}}$ varies more slowly compared with the sharp decrease for the spin-1 case, where increasing $u$ leads to a more rapidly increasing dependence of $I_{Q}/M^2$ on $M$. We might expect this on physical grounds, especially as the ground state of the spin-1 system begins to change substantially towards a transition to a spin-Mott regime near $u\simeq0.6J$.

\begin{figure}[tb]
\centering
\includegraphics[width=\columnwidth]{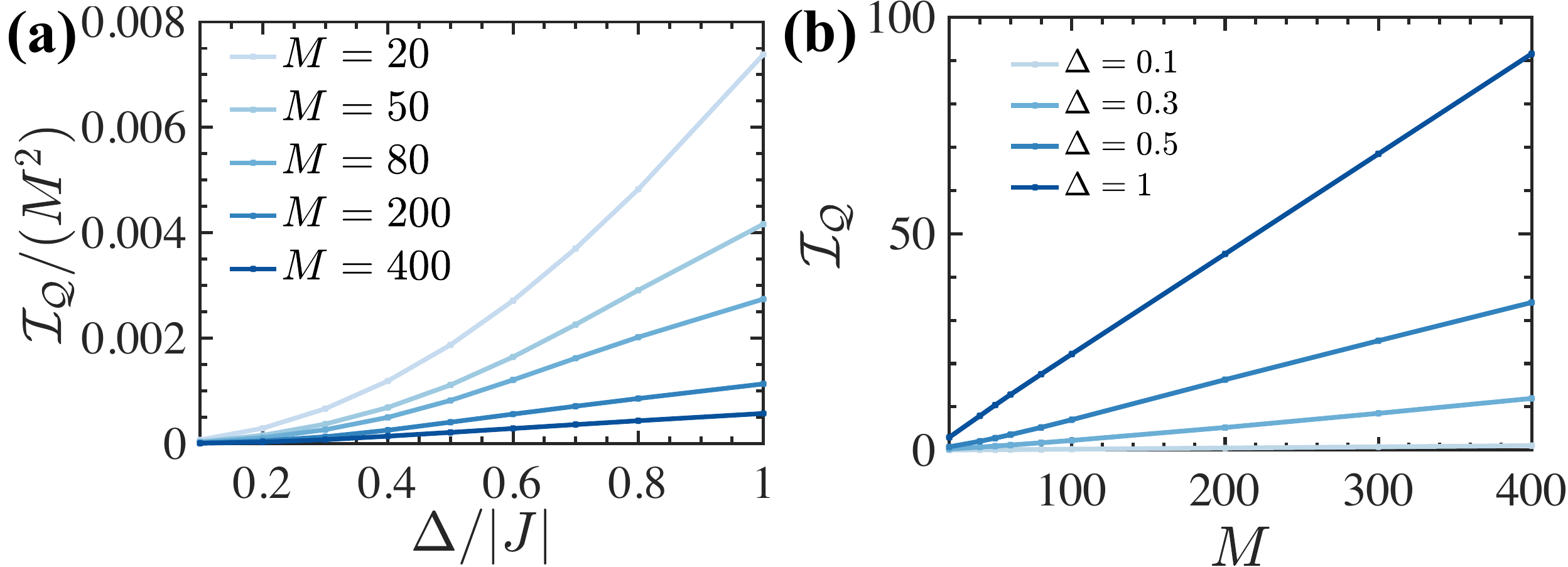}
\caption{(a) Quantum Fisher Information $\mathcal{I_{Q}}$ for the anti-ferromagnetic ground state of the Hamiltonian (spin-1/2 model) with $J<0$, vs. different values of the anisotropy $\Delta/J$. (b) $\mathcal{I_{Q}}$ versus system size, where we see the scaling is linear, thus the parameter $\alpha$ will be closer to the shot noise limit. The results are obtained via MPS methods with bond dimension $D\leq256$, with open boundary conditions.}
\label{fig:SP12_AFM}
\end{figure}

We also investigated the spin-1/2 case with the opposite sign of interactions, giving rise to an antiferromagnet (AFM) in the ground state, in Fig.~\ref{fig:SP12_AFM}. As in the spin-Mott, we see that the $\mathcal{I_{Q}}$ is very small. This is a demonstration that an antiferromagnet does not provide an opportunity for quantum enhanced measurement with this generator. This is unsurprising due to the alternating directions of spins. An open question is whether this could instead be adapted for enhanced measurement of processes involving generators that have a different spatial form locally. This might correspond, e.g., to a staggered spin direction (that could be used to sense a staggered magnetic field, or magnetic field gradient), which can thus be used to identify our antiferromagnetic phase~\cite{Gabbrielli2018}.

\section{Adiabatic State Preparation}
\label{sec:ASP}


Starting from a known occupation number of particles per site, we now investigate the dynamical preparation of specific magnetic states by appropriate time variation of external magnetic fields. The effective fields in different directions can be used in these systems to tune the interactions between atoms. Under various conditions we determine the fidelity of magnetically ordered states that can be engineered. This follows a variety of works focused on adiabatic state preparation with cold atoms in optical lattices and related systems \cite{Rabl2003,Rey2007,Ho2007,Soerensen2010,Schachenmayer2015,Trebst2006,Kantian2010,Lubasch2011}.

In the rest of the paper, we focus on the spin-1/2 model, as adiabatic state preparation (without the Fisher Information) was considered previously for the spin-1 model in \citep{Schachenmayer2015}. We concentrate on the XY-ferromagnetic regime in this manuscript, because of the large Fisher information demonstrated in the previous section.  Following other works \cite{Barmettler2010,Gammelmark_2013}, an alternative approach focuses on the adiabatic preparation and the study of many-body dynamics for the generation of anti-ferromagnetically ordered states. 

We investigate the spin-1/2 XXZ model with a small number of spins, represented by the Hamiltonian in \eqref{eq:Spin12-H}, with an extra effective external magnetic field. This arises from RF or Raman coupling of the spin states, treated in the rotating frame to give rise to the effective Hamiltonian
\begin{equation}\label{eq:XYFM_Hamiltonian_Ramp}
    \begin{split}
    \hat{\mathcal{H}}_{\text{SP1/2}}'= & - J{\sum_{\braket{i,j}}}({\hat{s}^x_i}{\hat{s}^x_j}+{\hat{s}^y_i}{\hat{s}^y_j})-(J-\Delta){\sum_{\braket{i,j}}}{\hat{s}^z_i}{\hat{s}^z_j} \\
    & -\Omega\sum_i{\hat{s}^x_i}.
    \end{split}
\end{equation}
We note that we should have $\Omega\ll U$ in order to maintain the Mott Insulator regime and the validity of perturbation theory to derive the spin models. However, it is still straight-forward to achieve $\Omega \gg J$, as $J=\zeta^2/U\ll \zeta \ll U$.

The adiabatic ramp we choose consists of two parts. To aid the preparation of a simple, high-fidelity initial state, we start with all spins in a superposition of $$\prod_i \frac{\ket{\uparrow}_i+\ket{\downarrow}_i}{\sqrt 2},$$ with a large coefficient of $\Omega/J$ (magnetic field in the $x$ direction - in practise in the experiment representing a strong coupling field between the two spin states). We reduce $\Omega/J$ to zero while fixing the anisotropy, $\Delta/J$. This is then reduced in a second ramp towards values close to zero, near the isotropic point where the Quantum Fisher information is large.

The first of these two ramps is assigned a time $T_{1}$. Below we will show that to achieve high-fidelity states in this ramp within reasonable experimental timescales, we will need to optimise the form of this ramp. For the second ramp, we choose a linear ramp in a time $T_{2}J$. This could also be optimised, but here we find that reasonable experimental timescales can be reached without doing so for the system sizes we calculate.  The total time for the protocol is $T=T_{1}+T_{2}$, and below we analyse each of the two ramps separately. 

\subsection{Optimizing the Ramp 1: $\Omega\to{0}$}
\label{sec:Optimal Control}





We find that this ramp can lead to exceedingly long preparation times for standard linear, exponential, or for ramps fitted to the energy gap (especially for small $\Omega/J$). Therefore, to make this ramp experimentally feasible, we resort to either shortcuts to adiabaticity \cite{del_Campo_2012,Campbell2017,del_Campo_2019}, or general optimal control techniques, which have been applied in chemistry and physics for a number of years \cite{Werschnik_2007,Brif_2010,Caneva2011}. In recent years, there has been substantial work on using these techniques to optimize state preparation in many-body quantum systems \cite{Doria2011,Glaser2015,van-Frank2016}, including speeding up adiabatic preparation in strongly interacting systems \cite{Brif_2014}. They can also be applied specifically to many-body dissipative dynamics \cite{Koch_2016}. 

Here we show that the very simplest form of this can make the ramps experimentally feasible with no difficult features (in practice this could also be optimized directly in the laboratory). Specifically, we chose to parameterize our time-dependent ramp as $\Omega(t) = \Omega_0 g(t)$ where $\Omega_0$ is the initial large field $\Omega_0=10J$, and
\begin{equation}\label{eq:Opt_Functions}
g(t)=\sum_{l}C_{l}f_{l}(t,\Theta_{l}).
\end{equation}
Here, $\Theta_{l}$ is an optimisation parameter and $f_{l}$ are the optimal control functions, where we fixed the duration of the evolution, i.e. the length of the ramp. We chose the rescaled Legendre polynomials for the control functions, providing an orthonormal basis over the interval $[-1,1]$.
We use a simple basis of 10 Legendre polynomials and the nonlinear optimization function \textit{fmincon} in \textit{MATLAB}. 

The figure of merit with which we quantify success in this case is the fidelity between the ground state (target state) of $\hat{\mathcal{H}}_{\text{SP1/2}}'$, or $\ket{\psi_{\rm{targ1}}}$, and the final state of the evolution $\ket{\psi(T_1)}$ at $t=T_1$, defined as:
\begin{equation}\label{fidelity}
F= {\vert\braket{\psi(T_1)|\psi_{\rm{targ1}}}\vert}^2.
\end{equation}

To provide reasonable computation times for larger system sizes, we choose a threshold where we search for a ramp that will give a final state fidelity $F\geq0.9$. This is already high (given that it is an operation that will only be applied once to the state), but we note that achieving even much higher fidelities is likely to be more straight-forward with more advanced optimal control techniques.

\begin{figure}[tb]
\centering
\includegraphics[width=\columnwidth]{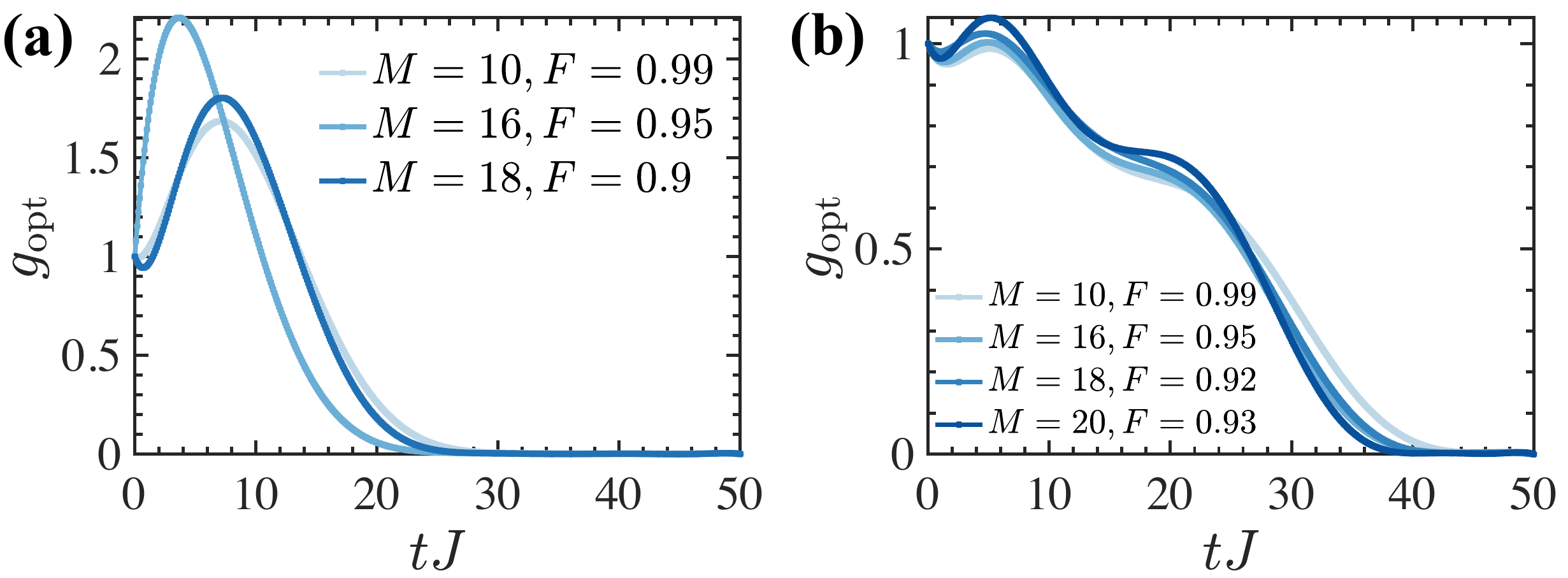}
\caption{Optimized ramps for the Hamiltonian in \eqref{eq:XYFM_Hamiltonian_Ramp}, where we ramp $\Omega\to 0$, for two different values of the anisotropy (a) $\Delta=0.8J$ and (b) $\Delta=2J$, and different system sizes $M$.}
\label{fig:Opt_Ramp}
\end{figure}
    
We call the optimized ramp $g_{\rm{opt}}$, and the results are shown in Fig.~\ref{fig:Opt_Ramp} for different system sizes and two anisotropy values $\Delta/J$ (high values in order to attain higher fidelities with this protocol). Some excited states are populated during the faster evolution allowing the procedure to maintain higher fidelity with a shorter time. The cost of optimization is non-trivial, but not computationally prohibitive. The ramps we produce are generally smooth, suggesting they should be easy to reproduce in an experiment. Naturally, the final state is sensitive to the exact shape of the ramp - however, this can be optimized in situ in experiments, and given the smooth shape of the ramp, this should be a relatively robust procedure. 

\subsection{Ramp 2: $\Delta\to{0}$}
\label{sec:DeltaRamp}

We are interested in approaching the isotropic XY-ferromagnet, $\Delta/J\to 0$, as we know from section~\ref{sec:metrology} that is the regime close to the Heisenberg scaling. As the first ramp works better for larger $\Delta/J$, we then consider a second ramp, of the anisotropy $\Delta/J$ and we evaluate the fidelity and the QFI $\mathcal{I_{Q}}$ during the ramp.

We start with the ground state $\ket{\psi_0}$ of the Hamiltonian in \eqref{eq:Spin12-H} with a specific value of $\Delta/J$, prepared following the protocol described in the previous section. We then ramp the anisotropy linearly, as $ {\Delta(t)} = \Delta- {\beta}t $, from $ {\Delta}/J$ at $tJ=0$ to $\Delta_T/J$ at time $ t=T_{2}$, for different values of ${\beta}$, and different final values of the anisotropy $\Delta_T$.

The fidelity between the target state $\ket{\psi_{\rm{targ}}}$ (the ground state of the Hamiltonian with $\Delta =\Delta_T$) and the final state  of the evolution $\ket{\psi(T_{2})}$ will then be:
\begin{equation}
F_{\Delta_{T}}= {\vert\braket{\psi(T)|\psi_{\rm{targ}}}\vert}^2.
\end{equation}

\begin{figure}[tb]
    \centering
    \includegraphics[width=\columnwidth]{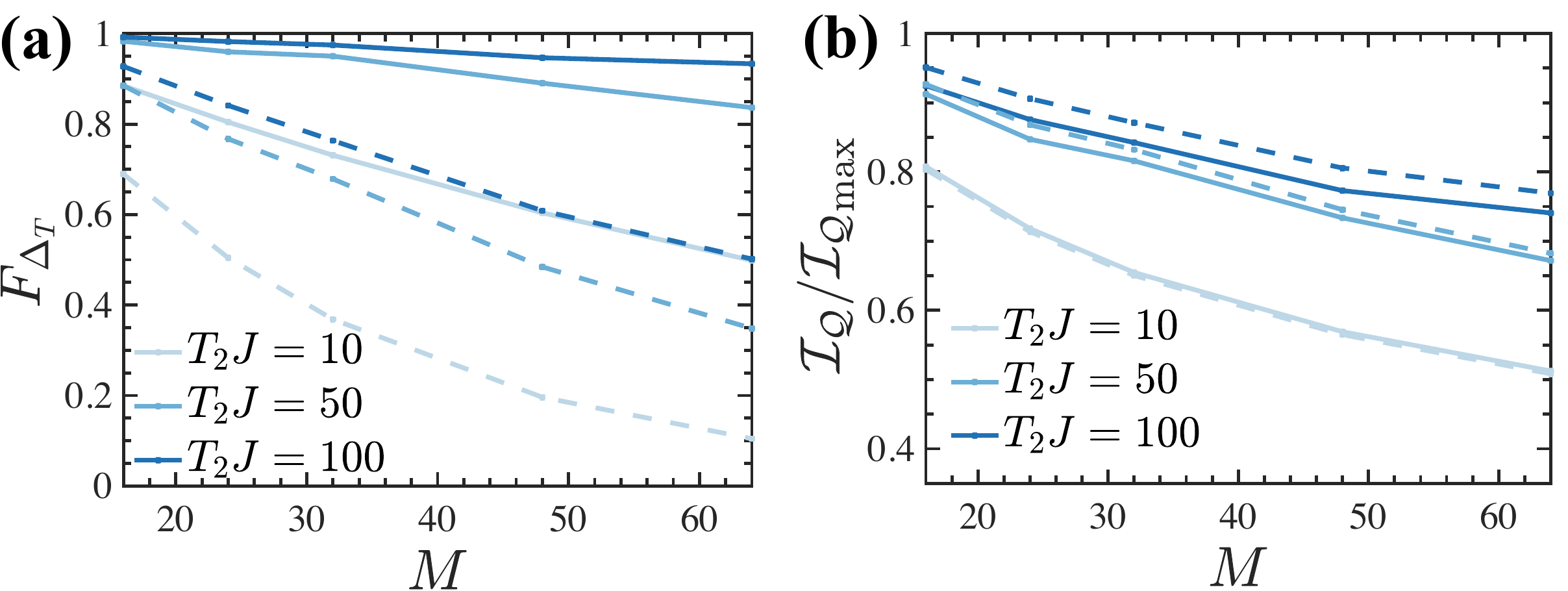}
    \caption{(a) Fidelities between the time-evolved state and the ground state at the end of the ramp in the anisotropy $\Delta/J$ for the spin-1/2 model, for $\Delta=0.8J$. We evaluate the ramps with two different final values of $\Delta_T/J$ (solid lines $\Delta_{T}=0.1J$, dashed lines $\Delta_{T}=0.01J$), for different system sizes $M$ and ramp times $T_{2}J$. We see that it is harder to target a smaller anisotropy $\Delta_T$, and that larger system sizes require longer ramps because of the smaller energy gaps in the spectrum of the Hamiltonian. 
    (b) Ratio of the Quantum Fisher Information to its maximum value $\mathcal{I_{Q}}/\mathcal{I_{Q}}_{\text{max}}$ at the end of the ramp in the anisotropy $\Delta/J$ for the spin-1/2 model, with $\Delta=0.8J$, different system sizes $M$ and different ramp times $TJ$.  The solid lines are for a final anisotropy value $\Delta_T=0.1J$ and the dashed lines for $\Delta_T=0.01J$. The decrease of the QFI with increasing system size is independent of the final value of $\Delta_T/J$, for a specific ramp time $T_{2}J$. These calculations were performed with bond dimension for the MPS calculations $D=128$, and open boundary conditions.}
    \label{fig:SP12_Ramp-M_B08}
\end{figure}

In Fig.~\ref{fig:SP12_Ramp-M_B08} we evaluate the final fidelity for different ramp times and different system sizes $M$, for two particular final values of the anisotropy $\Delta_T/J$. In agreement with the adiabatic theorem, the time scale required for the ramp to be adiabatic depends on the size of the system, since the energy gap will close as the size of the system increases. Furthermore, we identified how trying to reach lower final anisotropies becomes much harder with the system size. However, we show here that for typical experimental ramp times and system sizes we can still reach high fidelities up to a small final anisotropy value $\Delta_T$. We know the scaling is close to the Heisenberg limit with this $\Delta_T$ from Fig.~\ref{fig:QFI_GS}, which means we can still prepare these highly entangled states with high fidelity useful for precision measurements.

We also evaluate the QFI as in \eqref{eq:QFI} compared with its maximum value \eqref{eq:QFI_max}, and display it in Fig.~\ref{fig:SP12_Ramp-M_B08}. We identify that, contrary to the result with the fidelity, the final anisotropy has an insignificant role on the behaviour of $\mathcal{I_{Q}}$  with ramp time and system size. Thus, we understand that the procedure of preparing the ground state of the Hamiltonian \eqref{eq:XYFM_Hamiltonian_Ramp}, followed by a ramp in $\Delta/J$ to a final value of $\Delta_{T}=0.1J$, gives a practical means to generate an entangled state with a large QFI. 

We note that as in the ramps in $\Omega/J$, these ramps could be further optimized, and that ultimately there may also be a path in the 2D parameter space of $\Omega$ and $\Delta$ that might be found by the more sophisticated control methods mentioned above \cite{Werschnik_2007,Brif_2010,Caneva2011,Doria2011,Glaser2015,van-Frank2016} optimal in a given experiment. This is also influenced by heating in the system, which we will consider in the next section, and which could also be taken into account in optimal control \cite{Koch_2016}.

\section{Effects of dissipation on the adiabatic ramps}
\label{sec:Dissipation}

\begin{figure}[tb]
    \centering
    \includegraphics[width=\columnwidth]{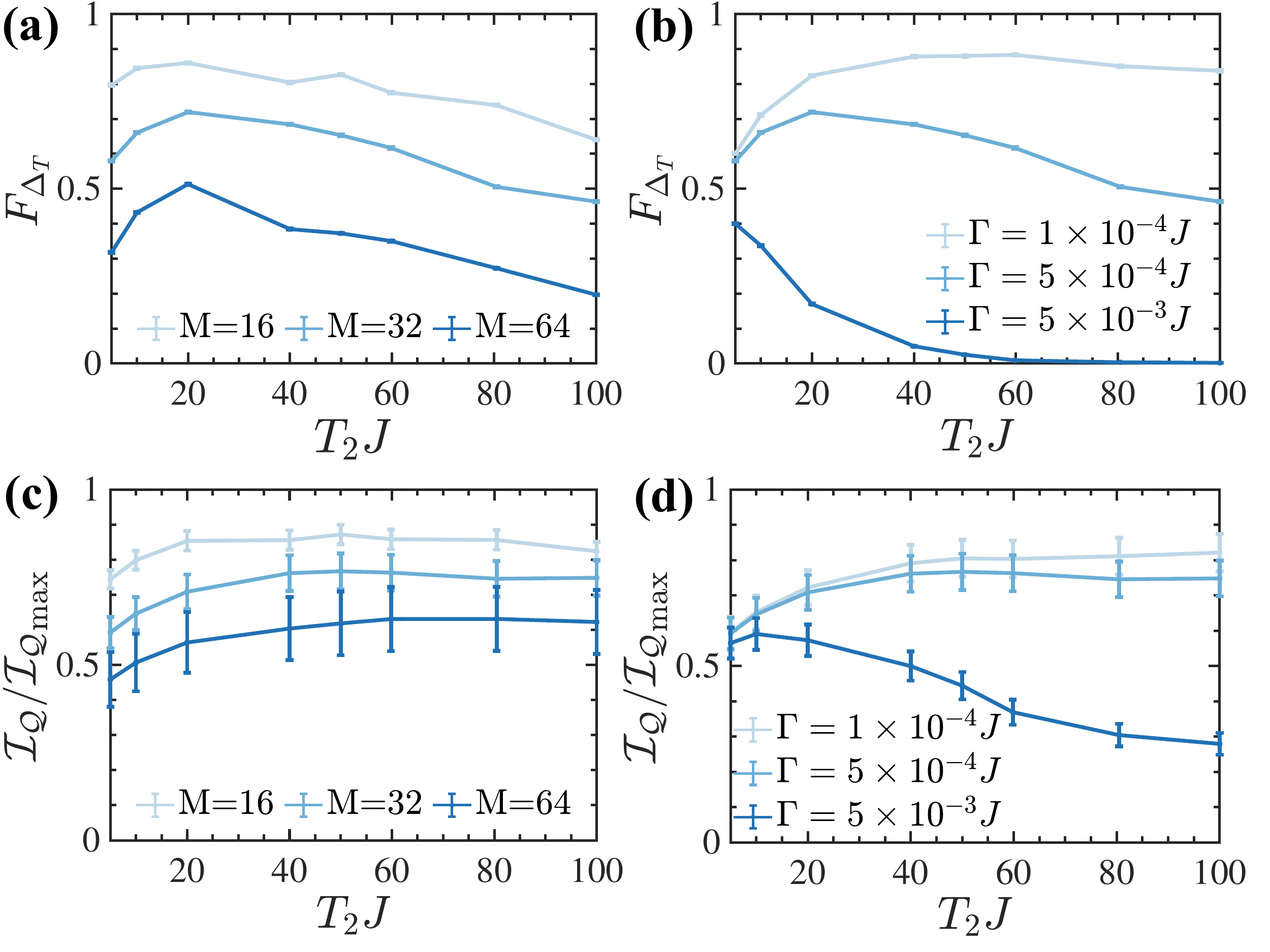}
    \caption{Averaged fidelities $F$ [panels (a-b)] and Quantum Fisher Information [panels (c-d)] at the end of the ramp in $\Delta$ including dissipation, for initial and final anisotropy $\Delta=0.8J$ and $\Delta_{T}=0.1J$, respectively.
    (a) Averaged fidelities for different system sizes, ramp times, and a value of the dissipation $\Gamma=5\times10^{-4}J$. (b) Averaged fidelities for different dissipation rates and ramp times, for a system of size $M=32$. 
    (c) $\mathcal{I_{Q}}/\mathcal{I_{Q}}_{\text{max}}$ shown for different system sizes, ramp times, and a value of the dissipation $\Gamma=5\times10^{-4}J$. (d) $\mathcal{I_{Q}}/\mathcal{I_{Q}}_{\text{max}}$ shown for different dissipation rates and ramp times, for a system of size $M=32$. These calculations were performed with 200 trajectories (statistical error bars are shown on the plots) and a bond dimension for the MPS calculations $D=128$, with open boundary conditions.}
    \label{fig:Dis_Ramp}
\end{figure}

One of the challenges in using adiabatic state preparation in open environments is the trade-off between the speed of the ramp and the energetic cost of natural heating. The most fundamental form for atoms trapped in optical potentials is spontaneous emissions, but depending on the experimental setup, this can also come from field fluctuations, and noise on the amplitude of the optical lattice \cite{McKay2011}. 

We model the most destructive source of heating, from spin dependent local fluctuations, or spin-sensitive spontaneous emisssion events \cite{Pichler2010,Daley_Traj}. The resulting markovian master equation for our system in Lindblad form is 
\begin{equation}
\begin{split}
&\frac{\text{d}}{\text{d}t}{\hat{\rho}}=-\frac{\mi}{\hbar}[\hat{\mathcal{H}}_{\text{SP1/2}},\hat{\rho}]\\
&-\frac{\Gamma}{2}\sum_{i,\kappa=A,B}\bigg[\hat{C}_{i,\kappa}^{\dagger}\hat{C}_{i,\kappa}\hat{\rho}+\hat{\rho}\hat{C}_{i,\kappa}^{\dagger}\hat{C}_{i,\kappa}-2\hat{C}_{i,\kappa}{\hat{\rho}}\hat{C}^{\dagger}_{i,\kappa}\bigg],
\end{split}
\label{eq:MasterEq1}
\end{equation}
where $\Gamma$ is the dissipation rate for an individual spin, and the jump operators on site $i$ are
\begin{equation}
\begin{split}
& \hat{C}_{i,A}=\frac{\mathbb{1}+\hat{s}^z}{2}, \\
& \hat{C}_{i,B}=\frac{\mathbb{1}-\hat{s}^z}{2},
\end{split}
\end{equation}
for species $A$ and $B$, respectively.

In order to solve this master equation numerically we employ quantum trajectories techniques, which involve rewriting the master equation as a stochastic average over a number of individual trajectories, which can be evolved in time numerically as pure states \cite{Daley_Traj}. 

For general mixed states produced in our dissipative evolution, the quantum Fisher information can be computed from the eigenvalue decomposition of the full density matrix~\cite{Pezze2018}, $\hat\rho = \sum_\alpha \lambda_\alpha \ket{\lambda_\alpha}\bra{\lambda_\alpha}$ (with eigenvalues $\lambda_\alpha$ and eigenstates $\ket{\lambda_\alpha}$), as
\begin{align}
    \mathcal{I_{Q}} 
    = 2 \sum_{\substack{\alpha,\beta \\ \lambda_\alpha+\lambda_\beta > 0} } 
    \frac{(\lambda_\alpha-\lambda_\beta)^2}{\lambda_\alpha+\lambda_\beta}
    \left|
    \bra{\lambda_\alpha} \hat J_x \ket{\lambda_\beta}
    \right|^2.
    \label{eq:QFI_mixed_states}
\end{align}
It is important to note that the evolution preserves the total magnetization $\hat J_z$, since both $[\hat{\mathcal{H}}_{\text{SP1/2}}, \hat J_z] = 0$ and $[\hat C_{i,A/B}, \hat J_z] = 0$. Therefore, the eigenstates of the density matrix remain eigenstates of $\hat J_z $, $\hat J_z \ket{\lambda_\alpha} = z_\alpha \ket{\lambda_\alpha}$. Since the initial pure state $\ket{\psi_0}$ is an eigenstate of $\hat J_z$ with zero magnetization, our final density matrix decomposition only contains eigenstates with $z_\alpha = 0$, and $\lambda_\gamma =0$ for corresponding eigenstates $\ket{\lambda_\gamma}$ with $z_\gamma \neq 0$. Therefore, $\bra{\lambda_\alpha} \hat J_x \ket{\lambda_\beta} = 0$ for states with $z_\alpha = z_\beta$. This implies that for our specific problem, Eq.~\eqref{eq:QFI_mixed_states} simplifies to
\begin{align}
    \mathcal{I_{Q}} 
    = 4 \sum_{\alpha, \beta} \lambda_\alpha
    \left|
    \bra{\lambda_\alpha} \hat J_x \ket{\lambda_\beta}
    \right|^2
    = 4 \text{tr}\left( \hat J_x^2 \hat \rho \right)
    \label{eq:QFI_mixed_state_zconserve}.
\end{align}
Here, we exploited the fact that for all non-zero terms in the sum of Eq.~\eqref{eq:QFI_mixed_states} either $\lambda_\alpha = 0$ or $\lambda_\beta = 0$. Though they are generally different, Eq.~\eqref{eq:QFI_mixed_state_zconserve} implies that for our scenario, the mixed state QFI is simply equivalent to the variance $4 \Delta \hat J_x$ as defined in Eq.~\ref{eq:QFIa}, but with mixed-state expectation values (Note that we always have $\langle \hat J_x \rangle= \text{tr}(\hat J_x \hat \rho) = 0 $). This also  means that the QFI is here independent on the decomposition chosen for $\hat \rho$ and we can use the statistic average of the quantum trajectories produced by our evolution method.

In Fig.~\ref{fig:Dis_Ramp} we plot both the fidelity and the QFI (calculated as before) at the end of the ramp for different ramp times and different values of the dissipation $\Gamma/J$, using realistic orders of magnitude for current experiments. The average number of jumps we get per trajectory for $M=16$ are $\approx0.07,0.36,3.61$ at $\Gamma=1\times10^{-4}J,5\times10^{-4}J,5\times10^{-3}J$, respectively, and these values duplicate when doubling the system size.
We show how there is a trade-off between using slow ramps to improve adiabaticity and using faster ones to avoid dissipation, which is especially visible from the peaks as a function of total ramp time in Fig.~\ref{fig:Dis_Ramp}(a,b). For large heating rates the final fidelities become very small. It is important to note that the effects of dissipation are not as strongly visible in the QFI, which remains large, and generally has maximal values at significantly longer ramp times than the fidelity. This implies that for practical applications, the generation of useful entanglement should in general be more robust to heating than would be implied by the state fidelity. This is a similar conclusion to that reached for certain types of correlation functions in ground states produced by adiabatic ramps in the presence of dissipation \cite{Schachenmayer2015}.

\section{Summary and Outlook}
\label{sec:Summary}

We have explored state preparation using adiabatic (and also optimized) ramps of parameters for two-component bosonic atoms in an optical lattice. Near the isotropic points of the corresponding spin-1/2 and spin-1 models, we find states with strong entanglement, reflected in the Quantum Fisher Information. We show that these points can be approached robustly for realistic experimental size and time scales, also in the presence of decoherence. 

This raises a more general question about optimal preparation of these states in experiments, and whether more sophisticated optimal control techniques can be used, also potentially optimizing ramps while accounting for dissipation \cite{Koch_2016}, to produce robust useful states in experiments. This also provides an opportunity to determine how the shape of the ramp should depend on the many-body physics near the isotropic point in the model. In the future, it would be interesting to look at a broader range of similar models, determining similar points where the entanglement is large and analyzing the practicalities of adiabatic state preparation in those cases \cite{Frerot2018,PhysRevX.8.021022,Sinatra2018,YE2016151,Liu_2013}.

One particularly interesting further point that comes out of our results is that the Quantum Fisher Information of the target state is more robust than the Fidelity against the dephasing we treat here. This is potentially interesting for applications in quantum enhanced metrology, and it opens an important new question on what type of measurement protocol should be engineered in experiments to make optimal use of the state after decoherence. This may also be a challenging direction to explore in the future to develop optimal control techniques. 

The data for this manuscript is available in open access
at \cite{Strath_Pure}.

\begin{acknowledgments}
    This work was supported by AFOSR MURI FA9550-14-1-0035. Work at the University of Strathclyde was supported by the EPSRC Programme Grant DesOEQ (EP/P009565/1), by the EOARD via AFOSR grant number FA9550-18-1-0064, and by AFOSR MURI FA9550-14-1-0035.
    Numerical calculations here utilized the ARCHIEWeSt High Performance Computer.
     W.K. receives support from the NSF through the Center for Ultracold Atoms and Award No. 1506369, from ARO-MURI NonEquilibrium Many-Body Dynamics (Grant No. W911NF14-1-0003), from AFOSR-MURI Quantum Phases of Matter (Grant No. FA955014-10035), from ONR (Grant No. N00014-17-1-2253), and from a Vannevar-Bush Faculty Fellowship. 
    J.S. is supported by the French National Research Agency (ANR) through the Programme d\textsc{\char13}Investissement d\textsc{\char13}Avenir under contract ANR- 11-LABX-0058 NIE within the Investissement d\textsc{\char13}Avenir program ANR-10-IDEX-0002-02.
    M.L.C. warmly thanks the University of Strathclyde for hospitality, while part of this work has been conceived, and acknowledges support from the project $\rm{PRA\_2018\_34}$ (``ANISE'') from the University of Pisa.
    M.L.C. and A.D.~thank KITP for hospitality during this work, supported by the National Science Foundation under Grant No. NSF PHY-1748958.
\end{acknowledgments}

\bibliographystyle{apsrev4-1}
\bibliography{Paper_MIT2}

\end{document}